\documentstyle[pra,aps]{revtex}
\begin{document}
\draft
\title{A Bethe lattice representation for sandpiles}
\author{Oscar Sotolongo-Costa$^{1,2}$, Alexei Vazquez$^1$and J.C. Antoranz$^2\thanks{%
Author whom all correspondence should be addressed}$}
\address{$^1$ Dpto F\'{\i}sica Te\'orica, Fac. F\'{\i}sica, Universidad de La Habana, Habana\\
10400, Cuba.\\
$^{2}$LCTDI, Fac. Ciencias, UNED, Madrid 28080, Spain.}
\maketitle

\begin{abstract}
Avalanches in sandpiles are represented throughout a process of percolation
in a Bethe lattice with a feedback mechanism. The results indicate that the
frequency spectrum and probability distribution of avalanches resemble more
to experimental results than other models using cellular automata
simulations. Apparent discrepancies between experiments are reconciled.
Critical behavior is here expressed throughout the critical properties of
percolation phenomena.
\end{abstract}

\pacs{PACS numbers: 64.60 Fr, 64.90 +b, 84.90 +n}

\section*{Introduction}

The idea of self-organized criticality (SOC) proposed by Bak, Tang and
Wiesenfeld \cite{bak1,bak2} triggered a lot of experimental as well as
theoretical work on relaxation processes in granular materials. Sandpiles
seem to be the simplest systems to test self-organized behavior. The
importance of its study comes from the fact that SOC has been suggested as a
possible explanation for the power law behavior seen in many systems:
earthquakes,\cite{elgazar} mass distribution in the Universe, star flickers,
etc.\cite{mandelbrot}

Experiments on sandpiles were designed and performed in \cite{jae,held}. In 
\cite{jae}, avalanche sizes were recorded in rotating drum experiments,
finding that avalanches, instead of being distributed over all sizes obeying
a power law distribution as predicted in \cite{bak1,bak2}, occurred quite
regularly in size and time, in an almost periodic pattern (See also\cite
{nagel,jae2}).

In \cite{held}, mass fluctuations in an evolving sandpile were studied,
showing that for small enough sandpiles, the observed mass fluctuations are
scale invariant, and probability distribution of avalanches shows finite
size scaling whereas large sandpiles do not. In this experiment, small
sandpiles seemed to exhibit SOC. Besides, an apparent disagreement has
emerged between the results reported in \cite{jae} and \cite{held} but, as
we will show in this paper, these results are essentially the same.

Though many other theoretical and experimental works were performed 
\cite{kadanoff,hwa,evesque,pelayo,linz,bouchaud,bau,ben,sen,alonso,elperin,kunt,mehta,soto,zap,stan,sornette,alstrom,vaz,malthe}, 
some of the later proposed models were devoted to
the problem of SOC in a more general fashion than the sole application to
sandpiles 
(e.g. \cite{kadanoff,hwa,pelayo,ben}). Some others fix their
attention in models for which particular mechanisms of interaction seem to
be relevant \cite{linz,bouchaud,bau,alonso,mehta,malthe}.

In the present work we propose a representation of the avalanche process in
sandpiles as a percolation in a Bethe lattice, capturing the essential
features of the avalanche phenomenon and simultaneously taking into account
the nature of the sandpile, in order to reproduce the experimental results.

The image of an avalanche as an initial object that consecutively drags
another resembles a branching process for which the Bethe lattice
representation seems to be natural. This branching process was proposed in 
\cite{soto}, showing good possibilities to describe the change of behavior
of fragment size distribution in fragmentation phenomena. Another branching
process representation was proposed in \cite{pelayo,alstrom} in an attempt
to obtain analytical solutions for avalanche processes, and in \cite
{zap,stan}, a self-organizing branching process was proposed with a closer
relation with properties of physical systems. There, a feedback mechanism is
introduced, but the branchig structure is not related to the physical nature
of the system.

The introduction of a Bethe lattice representation, as will be seen, has the
advantage of its generality because the process of dragging is characterized
by a {\em {drag probability}} $p$ for each one of the particles forming the
nodes of the lattice. The nature of the system is taken into account through
the relation of $p$ with the parameters characterizing the self- organizing
characteristics of the system, i.e. the slope angle $\theta $ and the size
(number of grains) of the pile, $N$. This viewpoint is similar to that of 
\cite{zap,stan} but with a much closer relation with the experiments and the
physical nature of the studied systems, i.e., sandpiles. The main features
obtained there can be reproduced with this representation, but we will focus
on the polemics related with experimental data \cite{nagel}.

In section 2 we describe the representation of the avalanches and expose the
relation between the drag probability and sandpile characteristics.

In section 3, the results of simulations with this representation are
exposed and compared with experimental results. Section 4 is devoted to the
conclusions.

\section*{Representing the sandpile and the avalanches}

The phenomena for which sandpiles seem to be the simplest paradigm should
manifest themselves in a representation independent of the avalanche
mechanisms. A representation through a Bethe lattice resembles well the
process, and to reveal the behavior of the phenomenon, only the essential
characteristics should be involved.

Let us represent the avalanche as a cascade process in the Bethe lattice as
follows: Firstly, we start with a single node, which could represent in this
case a grain. With probability $p$, from this node will emerge $F$ new
nodes, representing that the initial particle has generated $F-1$ new
indistinguishable particles, which in principle will continue the avalanche.
This operation (generation of new identical particles with probability $p$ )
is applied to each node of this new group. In this case some of them will
continue the generation, some will not, and so on. The process is
represented in fig.1 for $F=3$. Empty nodes represent those in which the
process of percolation in this lattice will not progress, mimicking those
particles that do not follow the falling process in the cascade. This
representation seems to be natural for the process and does not appeal to
the nature of the forces between the grains in the sandpile, the nature of
the grains or even the nature of the avalanche.

Once the percolation process overcomes a given length (that of the border of
the sandpile), those nodes beyond the limit constitute the avalanche.
Counting the number of grains that just surpass the length of the pile is
equivalent to measuring the size of the avalanche. Knowing the number of
grains in the pile, it is possible to register the mass variations in the
pile with time, $M(t)$. The percolation process stops once the limit is
surpassed; the pile {\sl reorganizes itself }with the new number of grains,
(i.e. a new slope is calculated with the remaining grains) and again starts
growing until a new avalanche develops. This establishes a feedback
mechanism in this model, linking it with the experimental conditions.

We now relate the percolation probability with the parameters characterizing
the pile, namely, its slope and size, using the simplest representation. To
do this let us represent a conic sandpile of height $h$, radius $R$ and base
slope $\theta $. A small sphere of ''effective radius'' $r_0$ can
characterize the size of the grain of sand. The ratio of the volume of the
pile to that of the grain (including porosity effects in the value of $r_0$)
gives the number of grains in the pile: 
\[
N=\frac 14x^3\tan \theta 
\]
where $x=R/r_0$ is the ratio of the radius of the pile and the effective
radius of the grain.

The percolation probability is in this case translated to a {\em {drag
probability}} $p(\theta )$ of one grain to the next $F-1$ situated down the
slope. In this way, the interpretation of fig.1 is straightforward.

The dependence of $p$ with \ $\theta $ can be formulated {\sl a priori }%
taking into account that the drag probability should be an increasing
function of the slope. A good variable to describe this slope seems to be $%
\tan \theta $. On the other hand, the drag probability should be small for
angles less than the critical, and, once surpassed that angle, the
probability for an avalanche to take place must increase sharply. Let us
propose the following dependence:

\begin{equation}
p(\theta ,T)=\frac{\exp (\frac{\tan \theta -\tan \theta _c}T)}{\exp (\frac{%
\tan \theta -\tan \theta _c}T)+1}  \eqnum{1}
\end{equation}

where $\theta _c$ is the critical angle, related with the Coulomb's law $\mu
=\tan \theta _c$ ($\mu $ is the friction coefficient and here, for
simplicity, will be taken as unity), and $T$ is a parameter through which we
may control the sharpness of the variation just at $\theta =\theta _c$ and
can be used to include factors like granularity and vibrations\cite{vaz}.
Using the relation between the slope angle and the number of grains in the
pile, in this case (1) can be expressed as:

\begin{equation}
p(y,T)=\frac 1{1+\exp (\frac{1-y}T)}  \eqnum{2}
\end{equation}

where $y=N/N_c$. $N_c$ is the number of grains corresponding to $\theta _c$
. This is valid for small $T$ so that the variation of $p(y,T)$ is sharp
near $y=1$, i.e. $T<<1/4.$

Once the pile arrives to a size around $N_c$ the avalanches will be
noticeable and the slope will be readjusted after each avalanche. The
process of adding grains will again vary the value of $p$ up to values near $%
p_c$, -the value of the probability corresponding to $\theta _c$-, to
produce another avalanche, and so on. This constitutes a mechanism of
feedback, since the flux of sand tends to be kept constant because of the
concurrence of sand supply and avalanches. This has been recognized as
necessary \cite{sornette} to deal with critical behavior. Because of size
effects, avalanches will be registered for $p$ slightly less than $p_c$.

If, for a given value of $p$ , an avalanche develops, it will be counted if
the number of steps in the Bethe lattice surpasses the length of the slope
of the pile. The length of the slope, $L$, measured in units of the grain
diameter, is $g=x/2\sqrt{1-p}$ , and this is the threshold for an avalanche
to be registered. When the value $\ g$ is surpassed, those grains (nodes in
the Bethe lattice) belonging to that generation, are counted as the size of
the avalanche. This will permit to relate the results of the simulations
with the measured magnitudes, namely, the size of the avalanches, $S(t)$ 
\cite{jae} or the mass of the piles, $M(t)$ \cite{held}, as a function of
time. In this last case, the mass of the sandpile is represented as the
number of grains $N(t)$.

\section*{Results of the simulations}

Simulations were performed for a wide range of values of $x$, ranging from
10 to 500, using (2) for $T=0.1$ and a Bethe lattice with two branches. For
each value of $x$, more than $2^{16}*100$ realizations were performed.
Collected data were $N(t)$ and $S(t)$.

The temporal fluctuations of the mass $N(t)$, measured as the number of
grains in the pile, in units of $N_c$, are plotted in figure 2 (a), (b) and
(c) for $x=$ 50, 100 and 500. The unit of time used was that between two
consecutive events of adding grains. Avalanches are considered as
instantaneous. This behavior resembles that reported in \cite{held}. Note
that different time scales were used in fig. 2 (a), (b) and (c) for a better
illustration of the time variation for different sizes.

Figure 3 shows the probability of avalanche sizes $P(s)$ scaled as $x^{1.9}$%
, as in \cite{held}, vs $s$ normalized to $x^{0.95}$ for $x=$ 50 and 500,
showing a good finite size scaling $P(s)x^\beta $ vs $s/x^\nu $ with $\beta
=2\nu $ as in \cite{held}. The exponent $\nu =0.95$ was chosen as the best
fitting for all data with $s/x^\nu <2$. This representation reproduces the
finite size scaling for all sizes of sandpiles. The exponential falloff can
be verified. Larger piles show more dispersion for large avalanche sizes.
This can be explained noting that in the percolation process, large
percolation lengths correspond to probabilities near the critical one, where
critical behavior dominates and fluctuations are stronger.

Figure 4 (a) and (b) shows the power spectrum of $N(t)$ and $s(t)$
respectively for $x=$ 50, 100 and 500 exhibiting in (a) a clear $1/f^2$%
dependence, which also coincides with the results of \cite{held}. Power
spectrum in (a) reveals the same characteristics for all sandpiles, i.e.
dependence $1/f^\beta $ with $\beta \simeq 2$ , whereas in (b), the power
spectrum is clearly flat as in \cite{jae}.

Concerning the apparent disagreement between the experimental results in 
\cite{jae} and \cite{held}, it must be said that the process of measure in
both experiments is essentially different, since in \cite{held} the mass of
the pile is recorded as a function of time, whereas in \cite{jae} the
experiments record the variation of avalanche size as a function of time,
i.e., the magnitude that would correspond to the temporal variation of the
number of grains in the rotating cylinder, that means its time derivative.
This leads to a different power spectrum, so that if the power spectrum
obtained in \cite{held} is $1/f^2$, in this case its derivative should have
a ''flat'' spectrum, in correspondence with the results of \cite{jae}, and
there is no disagreement. In our case the spectrum is flat for high
frequencies because we are considering the avalanches as instantaneous, but
this is not an essential point. The simulations could be improved
introducing a finite time for avalanches. Though both teams have argued
about the differences of their experimental setups, we think that our
argument shows a very important difference.

The sizes of the avalanches $S(t)$ represent the value of the variation of
the function $N(t)$ in each jump, i.e., the derivative of that function, so
that the spectra in \cite{jae} and \cite{held}, though could be interpreted
as expressions of different behavior, are really intimately related.

The particular dependence of the drag probability with the slope is not of
great importance for the main results of this work. Dependences as $\sin
\theta $, $\sin ^{2}\theta $ and others can be used in the simulations
without significant changes in the results concerning probability
distribution of avalanches, power spectrum, etc. The main property required
is the increase of the drag probability with the slope. Also, the number of
terminals in the Bethe lattice is not important for the main conclusions.
This reveals the robustness of this phenomenon.

\section*{Conclusions}

A Bethe lattice representation linked with characteristics of sandpiles,
including a feedback mechanism, has been presented in the same direction
outlined in \cite{zap} with a new viewpoint, closely related with the
physical nature of the sandpile, which leads us to a closer link with the
experiments. The reproducibility of the experimental results lies on the
fact that the Bethe lattice representation unravels the physical nature of
the avalanche process in sandpiles and is able to be linked with the
geometrical properties of the system.

This representation keeps the same nature for all avalanches, irrespective
of pile size. In the avalanche process it seems to exist some kind of
transition, manifested in the change of behavior of the size distribution of
avalanches when the piles are large, reflected in the increase of
fluctuations in the region of large avalanches. This reflects that the
process of percolation in the Bethe lattice approaches the critical point so
that the second order nature of the phenomenon reveals itself in the
language of percolation phenomena. In this way, the description of
avalanches has been translated to the problem of percolation in a Bethe
lattice, and in this sense the phenomenon is critical. Thus, SOC, examined
with this viewpoint, is present in the organization of avalanches.

The proposed representation seems to reproduce the behavior of the
sandpiles, is very simple to instrument in a computer, and reveals an
essentially unique behavior in small and large sandpiles. Besides, it
contains the main characteristics of sandpiles in the sense of the increase
of avalanche probability by adding sand grains and a readjustment of the
slope after each avalanche. Oscillations \ of $p$ and $\theta $ near a
critical value are properties of this model as they are also in the
branching process model proposed in \cite{zap,stan}. Finite size scaling for
different sizes was obtained for the distribution of avalanches with good
reproduction of the experimental behavior.

\section*{acknowledgment}

This work was carried while one of the authors (O.S.) was visiting the
U.N.E.D. in Madrid, Spain. The financial support by the Vicerrectorado de
Centros Asociados of the U.N.E.D. is gratefully acknowledged. This work was
partially supported by the Direccion General de Investigaci\'{o}n
Cient\'{i}fica y T\'{e}cnica (DGICYT, Spain, Ministerio de Educacion y
Cultura), and by the Alma Mater prize, Havana University. We are greatly
indebted to R. Garcia Pelayo for his helpful suggestions and comments.

\begin{figure}[tbp]
\caption{Bethe lattice representation of the sandpile behavior.}
\label{fig1}
\end{figure}

\begin{figure}[tbp]
\caption{Fluctuations of the mass of the sandpile $N(t)/N_{c}$ for (a) $x=$
50, (b) $x=$ 100 and (c) $x=$ 500 obtained from simulations in a Bethe
lattice as described in the text. The behavior for other number of terminals
in the lattice, or another plausible dependence of $p(\theta )$ is
essentially the same.}
\label{fig2}
\end{figure}

\begin{figure}[tbp]
\caption{Avalanche probability scaled as $x^{1.9}$ vs. avalanche size $s$
normalized to $x^{.95}$ for $x=$ 50 (circles) and 500 (squares). The result
for this theoretical sandpile resemble those of [6] with exponential falloff.
 Large piles show larger dispersion for large avalanches.}
\label{fig3}
\end{figure}

\begin{figure}[tbp]
\caption{(a).- Power spectrum of mass fluctuations for different sizes of
the sandpiles $x=$ 50, 100 and 500 as indicated in the figure. The spectrum
is $1/f^2$ in agreement with [6]. then the corresponding spectrum for the
avalanches is flat as in [5]. (b).- Power spectrum of the avalanche sizes for
the same set of values as in (a).}
\label{fig4}
\end{figure}

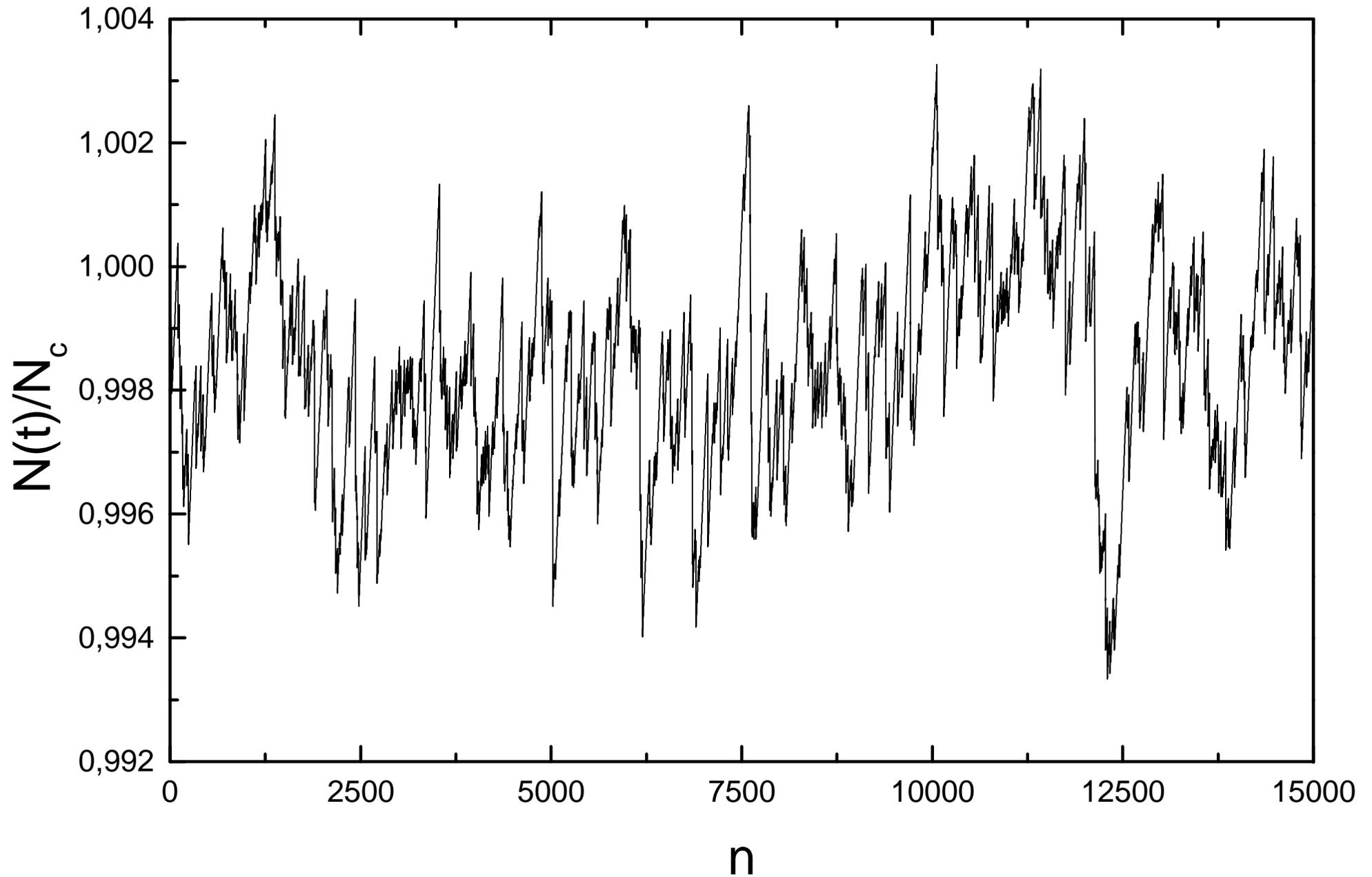
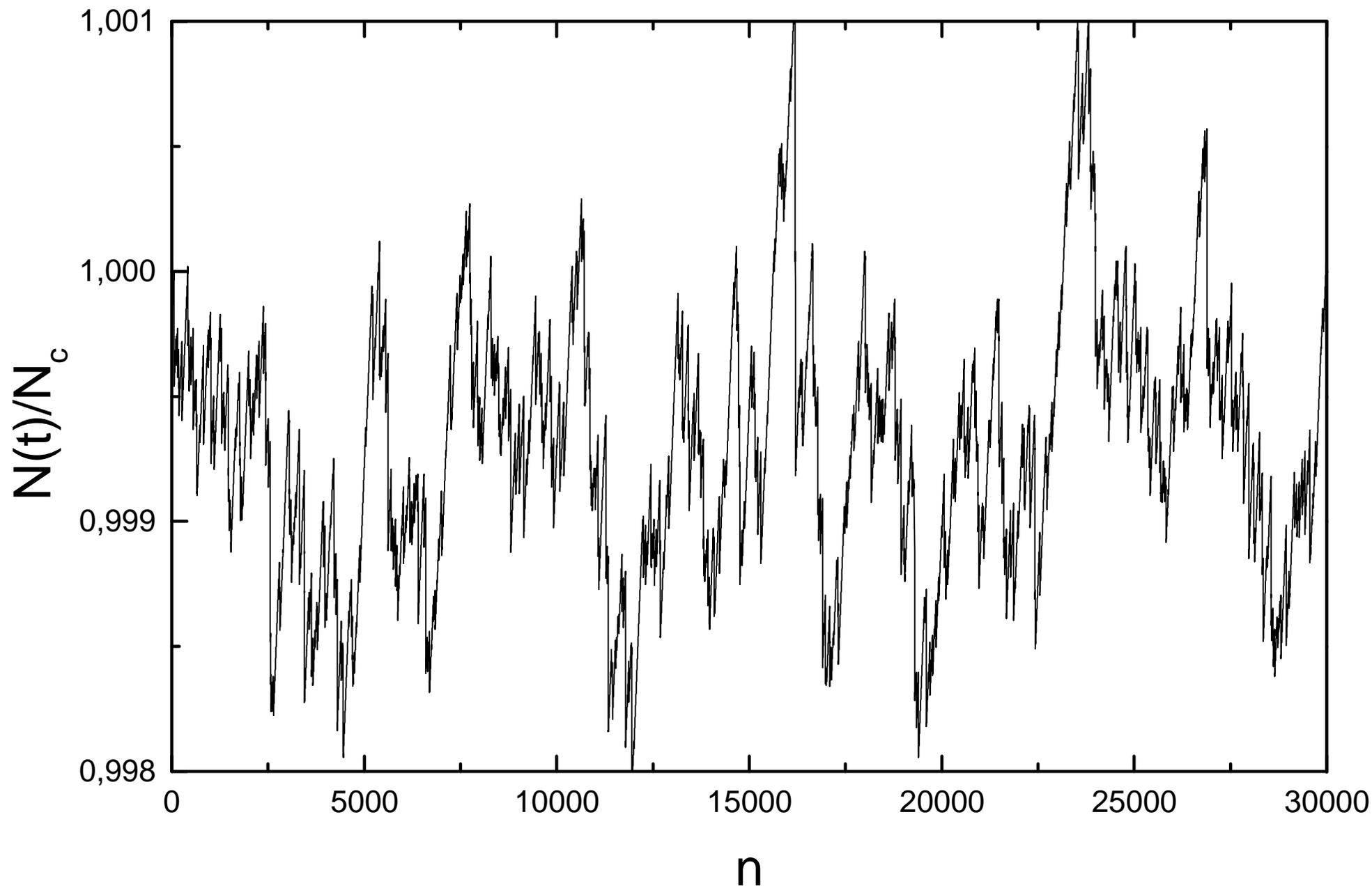
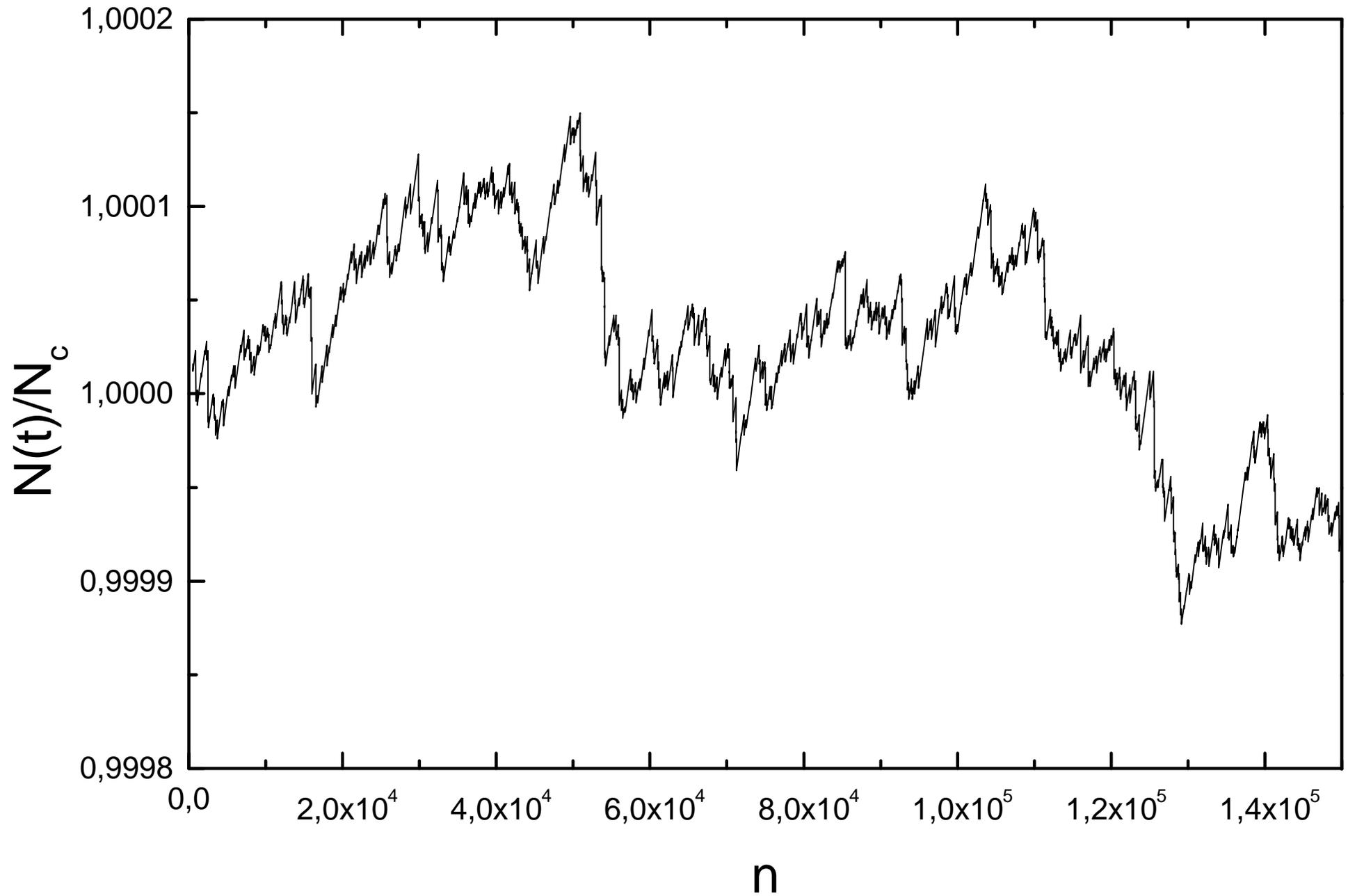
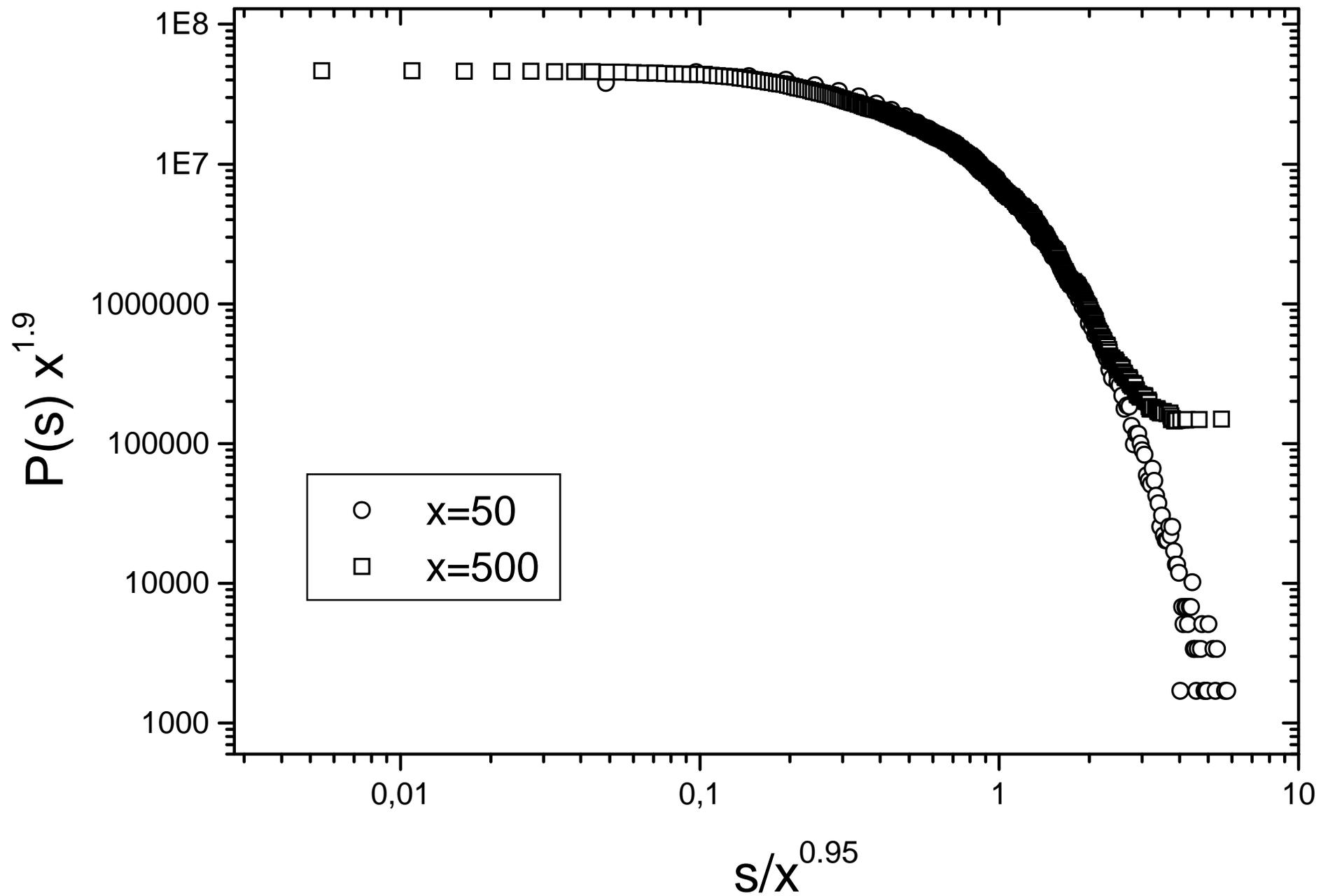
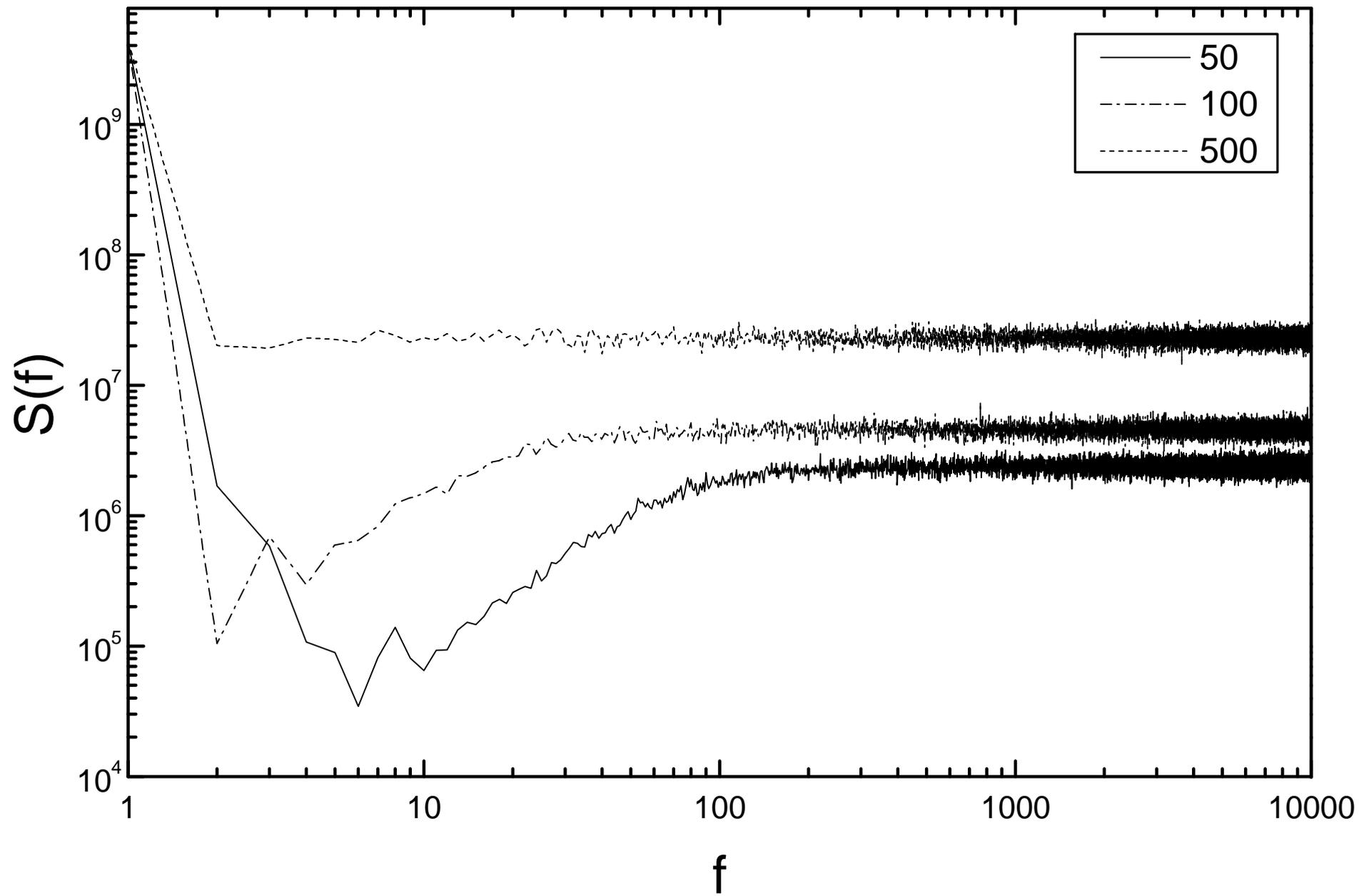
\end{document}